# Direct imaging of electron transfer and its influence on superconducting pairing at FeSe/SrTiO$_3$ interface

Weiwei Zhao[1]*, Mingda Li[2,3,4]*, Cui-Zu Chang[1,3]*, Jue Jiang[1], Lijun Wu[4], Chaoxing Liu[1], Yimei Zhu[4], Jagadeesh S. Moodera[3,5] and Moses H. W. Chan[1]

[1]The Center for Nanoscale Science and Department of Physics, The Pennsylvania State University, University Park, PA 16802-6300, USA

[2]Department of Mechanical Engineering, Massachusetts Institute of Technology, Cambridge, MA 02139, USA

[3]Francis Bitter Magnet Lab and Plasma Science and Fusion Center, Massachusetts Institute of Technology, Cambridge, MA 02139, USA

[4]Condensed Matter Physics and Materials Science Department, Brookhaven National Laboratory, Upton, New York 11973, USA

[5]Department of Physics, Massachusetts Institute of Technology, Cambridge, MA 02139, USA

*These authors contributed equally to this work.

Correspondence to: mhc2@psu.edu (M.H.W.C.); zhu@bnl.gov (Y.Z.); czchang@mit.edu (C.Z.C.).

**The exact mechanism responsible for the tenfold enhancement of superconducting transition temperature ($T_c$) in a monolayer iron selenide (FeSe) on SrTiO$_3$(STO) substrate over that of bulk FeSe, is an open issue. We present here a coordinated study of electrical transport and low temperature electron energy-loss spectroscopy (EELS) measurements on FeSe/STO films of various thicknesses. Our EELS mapping across the FeSe/STO interface shows direct evidence of band-**

**bending caused by electrons transferred from STO to FeSe layer. The transferred electrons were found to accumulate only within the first two atomic layers of FeSe films near the STO substrate. Our transport results found a positive backgate applied from STO is particularly effective in enhancing $T_c$ of the films while minimally changing the carrier density. We suggest that the positive backgate tends to 'pull' the transferred electrons in FeSe films closer to the interface and thus further enhances both their coupling to interfacial phonons and the electron-electron interaction within FeSe films, thus leading to a huge enhancement of $T_c$ in FeSe films.**

While bulk FeSe has a superconducting transition temperature $T_c$ of 9.4K[1], a scanning tunneling microscope (STM) study of one unit cell (UC) of FeSe film on SrTiO$_3$ (STO) substrate in 2012 found a superconducting gap of 20mV, suggesting a transition temperature $T_c$ of ~77K[2]. In spite of the ensuing intensive experimental[3-18] and theoretical studies[19-23], the exact mechanism responsible for the highly enhanced $T_c$ in this system is still lacking. An angle resolved photoemission spectroscopy (ARPES) experiment[6] found a replica band in a 1UC FeSe film on STO attributable to the strong coupling between electrons in the FeSe layer and an optical phonon mode of the underlying STO. This is consistent with the proposal that the high $T_c$ of the FeSe/STO system is due to stronger pairing of the electrons in the FeSe film enabled by the STO phonons[19]. A number of recent experiments investigated the effect of doping electrons into 3UC and thicker FeSe films grown on different substrates (STO, MgO or graphene) and also bulk-like FeSe flakes by depositing potassium[24,25] or liquid-gating layers[26,27] onto these films/flakes and enhancements of $T_c$ to near and above 40K were found. A recent STM study by Tang *et*

*al*[28] found the superconductivity in 1UC FeSe films grown on STO is continuously suppressed with increasing potassium coverage. Tang *et al* also found the deposition of potassium can induce superconducting gaps in 2UC FeSe/STO films previously not seen by STM[2] and ARPES[9] techniques. These electron-doping experiments led to the suggestion that STO plays a similar role as potassium in enhancing the $T_c$ of the 1UC FeSe/STO system, namely as an electron donor to the film. However, there is as yet no direct experimental observation on the spatial profile of such a charge transfer across the interface from STO to the FeSe film. The spatial profile not only provides a more precise picture of the transferred charges, it also clarifies the $T_c$ enhancement mechanisms in this system.

We report here a complementary Hall transport and atomic resolved electron energy-loss spectroscopy (EELS) study on the same 1UC, 8UC and 14 UC FeSe films on STO. Transport measurements were also carried out in a number of other 1UC and 2UC films. The high spatial resolution (~0.02nm) EELS mapping across the FeSe/STO interface at 10K shows clear and direct evidence of the band banding caused by electron charge transfer from STO into the FeSe films. As a consequence of band bending effect, the transferred electrons are accumulated within the first two FeSe UC from FeSe-STO interface, irrespective of the thickness of the FeSe film. Our transport measurements confirm that the $T_c$ of the 1UC films can be enhanced by thermal annealing which introduces electrons into the film[4,8] and also by gating from the back surface of STO with a positive potential[8]. Our systematic transport measurements allow us to conclude that backgating is particularly effective in enhancing the $T_c$ without changing significantly the carrier density of the films. This lends support to the model of electron-phonon coupling across interface as the mechanism for high $T_c$[6,18,19] since the positive potential tends to 'pulls' the interfacial electrons closer to the high Debye temperature STO phonon bath. We note that the electron-electron interaction effect[29,30] should also be strengthened by gating in such an interfacial two-dimensional electron gas (2DEG) as compared to bulk carriers due to the lowering of its dimension[31].

**Results**

**Superconducting FeSe/STO films.** The FeSe films were grown on $TiO_2$ terminated insulating STO(001) substrates (schematically shown in Fig.1a). Four 1UC-thick films were prepared by annealing at different temperatures for 2 hours after growth: S1(550℃), S1'(500℃), S1''(400℃), S1'''(330℃). Fig. 1e shows S1, the film annealed at 550℃ has the highest $T_c$. Three thicker FeSe films of 2UC, 8UC and 14UC were also annealed at 550℃ for 2 hours and are denoted as S2(2UC), S3(8UC) and S4(14UC). An additional 14UC FeTe layers were deposited on all samples as the capping layers[7] to prevent contamination for *ex-situ* transport and EELS measurements. The cross-sectional high-resolution scanning transmission electron microscope (STEM) images of the 1UC(S1), 8UC(S3) and 14UC(S4) show the interfaces of FeSe/STO and FeTe/FeSe are nearly free of defects (Figs. 1b-c). There is an interstitial layer between STO and FeSe films, as shown in Figs.1b and 1c. Similar structures have been reported[16,32] and the room-temperature EELS study on FeSe/STO system[16] identified the layer as two $TiO_x$ layers with increased oxygen vacancies.

The transport measurements were performed with the six-terminal Hall-bar geometry as depicted in Fig.1d. Indium foil was pressed onto the back of the 0.5mm thick STO substrates as the backgate electrodes. By applying $V_g$=+200V, the accumulated electron density at the interface in our sample configuration can be roughly estimated to be $1.84\times10^{13}$ cm$^{-2}$ at 30K. Figs.1f-i show the superconducting transitions for 1UC(S1), 2UC(S2) and 8UC(S3) films shifting to higher temperature with increasing backgate voltage $V_g$, while the 14UC(S4) film shows weak gate dependence. The $T_c$, defined as the temperature at which the resistance is reduced to half of the normal state resistance at 50K, changed from 21.8K at $V_g$=-200V to 27.0K at $V_g$=+200V (Fig.1f). The changes in $T_c$ for the 2UC and 8UC films were found to be higher, by 30% and 86% respectively, for the same change in gating voltage (Figs.1g-h). The $T_c$ found here for the 1UC film is substantially lower than that reported in STM[8], ARPES[4-6] and *in-situ* transport studies[13]. This can be explained by the fact that the FeTe capping layer introduces hole carriers into the FeSe film (Section IX of Supplementary Information).

**EELS measurement results.** The EELS mapping was performed across the FeSe/STO interface on 1UC(S1), 8UC(S3) and 14UC(S4) films at 300K and 10K after transport measurements on these films were completed. The core-loss EELS measures the energy difference between Fe $L_3$ ($2p_{3/2}$) and $L_2$ ($2p_{1/2}$) energy levels to the first unoccupied electronic states. The data were taken simultaneously in dual EELS mode at each spatial position. Figs. 2a-c show the core-loss EELS mapping with the energy level as the horizontal axis and the spatial position along the thickness direction as the vertical axis for the 1UC, 8UC and 14UC films. A blue-shift of the Fe's $L_3$ edge extending into the FeSe film from the STO interface is observed in all 3 samples at 10K (Figs.2d-f), but not seen at 300K (Section III of Supplementary Information). For the 8 and 14 UC films, the shifts smoothly diminish inside FeSe and disappears ~2UC away from the interface. The maximum magnitudes of the blue shift at the interface shown for the 8 and 14UC films are ~0.7±0.1eV (Fig.2e) and ~0.4±0.1eV (Fig.2f), respectively. For the 1UC film, the blue shift is clearly present (Fig.2d) but the magnitude is hard to estimate, due to the limited energy and spatial resolution resulting from signal delocalization and probe broadening.

The blue-shift for the Fe's $L_3$ edge indicates the possibility of electronic structure change in FeSe films near the STO substrate. To understand the origin of the energy shift, we first performed Green's function based EELS spectral simulation using the program FEFF[33-36]. The local chemical environment of the FeSe films will change when it is in direct proximity with STO surface. However, such effect can be excluded as the origin of the shift, since no energy shift is found in the main Fe's $L_3$ in the simulation, even for the Fe ions at the 1st UC closest to STO (Fig. 2g). The effect of the STO shows up as a shoulder in the spectrum on the high energy side of the main peak near 712eV. The shoulders found at the 2nd and 3rd UC are nearly identical. The strain effect can also be excluded since FeSe is experiencing tensile stress near STO[5], which results in a slight red shift instead of the observed blue shift (Section V of Supplementary Information).

We propose that the band bending at the interface induced by work function difference between the FeSe film and the STO substrate as a very reasonable explanation for the observed blue shift. The STO substrates used in experiments are electron-doped due to oxygen vacancies at the $TiO_x$ interstitial layer between STO and FeSe films[5,16,21]. The electrons reside in impurity bands are typically close to the bottom of the STO conduction band, while the Fermi level in FeSe is close to the top of the STO valence band[20], as schematically shown in Fig.2h. Consequently, the work function difference between FeSe and STO is close to the STO band gap (~3eV) and can lead to a charge transfer from STO to FeSe. The build-in electric field induced by this charge transfer yields a band bending across the FeSe/STO interface, as depicted in Fig. 2i. As a result of the band bending effect, the energy difference between the energy of Fe $L_3$ ($2p_{3/2}$) and $L_2$ ($2p_{1/2}$) levels near the interface and the Fermi energy is enlarged, giving rise to the blue-shift spectra of these core levels in the electron energy loss. The band bending picture also provides an explanation of the higher energy shift observed in the 8UC film than that in the 14UC film (~0.7 vs. ~0.4eV). Our Hall transport measurements (shown in Fig.3) show more hole-type carriers in 14UC film as compared to those in 8UC film, suggesting a lower Fermi energy $E_F$ in 14UC film than that in 8UC film. According to previous first-principles calculations on a FeSe film[22], the density of states increases when the Fermi energy is lowered deep into the valence band, as schematically depicted in Fig. 2j. Thus, we expect a smaller density of states at the Fermi energy for the thinner film and thus larger band bending with the assumption that same amount of electrons were transferred. In order to arrive at a more quantitative estimate, we make the following simplifications: (1) the transferred charges are assumed to be uniformly distributed in the FeSe layers with a thickness $d$ around 1~2UC (0.55~1.1nm) near the interface, (2) All the transferred electrons come from 2DEG on STO (001) surface[37,38] with a sheet carrier density $n$~0.5-1.5×10$^{14}$cm$^{-2}$ and (3) the dielectric constant of the FeSe film $\varepsilon_{FeSe}$ ~15[23]. These simplifications lead to the estimate of the voltage drop $V_B$ to be 0.1~1V according to the equation $V_B = \frac{en \cdot d}{2\varepsilon_{FeSe}}$, which is consistent with the observed energy shift (8UC: ~0.7eV, 14UC: ~0.4eV). This consistency buttresses the conclusion that the blue shifts observed in the EELS core-loss mapping are direct evidence of electron transfer from STO to the FeSe films.

**Hall transport results.** In addition to longitudinal transport and EELS measurements, we also systematically performed Hall transport measurements under different $\mu_0 H$, $T$ and $V_g$ on the three films studied by EELS, i.e., 1UC(S1), 8UC(S3) and 14UC(S4) and also 2UC (S2) and the three 1UC films (S'1,S1",S1"') annealed at lower temperatures (Fig.3 and Section VII of Supplementary Information). The Hall resistances $R_{yx}$ were measured as a function of magnetic field $\mu_0 H$ from -8 to 8T at different fixed $T$ ranging from 300K down to 30K. Linear dependence of $R_{yx}$ on $\mu_0 H$ is found for all samples. The $T$ dependences of the Hall coefficient $R_H = R_{yx}/\mu_0 H$ for all seven samples are shown in Figs. 3c and 3e. $R_H$ changes sign from positive to negative upon cooling from 300K to ~140K for the 1UC(S1), 2UC(S2), 8UC(S3) films. This behavior has been observed in many multiband materials in the presence of both electron-type and hole-type carries, such as NbSe$_2$[39]. Below 50K, $R_H$ of the 8UC film turns positive again. $R_H$ for the 14UC film stays positive for all temperatures with a rapid increase in value below 100K. This behavior indicates that the hole densities $n_h$ in both 8UC and 14UC FeSe films are enhanced substantially at low $T$ (Section VII of the Supplementary Information). This hole carrier dominated behavior for the thicker film is similar to that found for bulk FeSe with low $T_c$. Fig. 3e show that annealing the 1UC film at higher temperatures makes the 1UC films (S1, S1', S1") to be more electron carrier dominant at low $T$ and as shown in Fig. 1e, with higher $T_c$[3,7].

The gate dependence of $R_H$ for all seven samples has also been studied. $R_{yx}$ vs. $\mu_0 H$ at 30K for the 1UC film under different $V_g$ are shown in Fig.3b and the data for other samples are shown in Section VII of the Supplementary Information. Fig. 3d shows that the effect on $R_H$ due to a change in backgate voltage from -200 to +200V is an order of magnitude smaller than the difference in $R_H$ of films of different thicknesses. Similarly, weak dependences of $R_H$ on gating are also seen for the four 1UC samples subjected to different annealing procedures (Fig. 3f).

Although backgating is not effective in changing $R_H$ and hence the carrier density, it leads to a significant change in $T_c$ (Fig. 1). Fig.4 summarizes the effect of annealing, film thickness variation and backgate voltage on $T_c$ and the Hall coefficient $R_H$ of the seven samples, and show the gating effect for $T_c$ enhancement is most effective compared with the annealing and varying thickness.

**Discussion**

In this work we established a coherent picture on the issue of charge transfer in the FeSe/STO system. The EELS mapping of the FeSe/STO films provides a direct evidence of band bending induced by charge transfer across the interface from STO to the FeSe films. The transferred charges, irrespective of the thickness of the FeSe films, are found to accumulate as a 2DEG within the first two FeSe layers next to the STO substrate. An interstitial layer is observed between the FeSe film and the STO substrate and was identified as two TiO$_x$ layers with increased oxygen vacancies[16], this layer could be the source of the transferred charges.

Our transport measurements showed that backgating from STO is particularly effective in enhancing the $T_c$ of thin FeSe films without significant influence on the total carrier density. Combining with the observations in EELS measurements, this suggests a consistent physical picture that the enhancement of superconductivity in FeSe films on STO is due to the 2DEG transferred from STO and confined near the interface. Since FeSe film is a metal, the accumulated charges by the gating effect is much smaller compared to hole-carrier density in the system. The electric field generated by the backgate will be screened by the accumulated charges at the STO/FeSe interface and its effect is limited within the screening length. Thus, the backgate is not efficient in tuning the total carriers in FeSe films and its main influence is to 'pull" electrons in FeSe films closer to the STO interface. At the FeSe/STO interface, electrons can benefit from high Debye temperature STO phonon bath and thus the electron-phonon coupling for superconductivity is enhanced[6,18,19,40]. On the other hand, lowering the dimension by forming 2DEG at the interface[31] and separating electrons and holes in different layers can also enhance the electron-electron interaction, thus strengthen the mechanism of electron-electron interaction for superconductivity, similar to that in strongly correlated electron systems of the ordinary electron-type iron-based superconductors[29,30]. Thus, by combining the EELS mappings and transport results, we arrive at the conclusion that both electron coupling to STO phonons and electron-electron interaction within the transferred charges could be responsible for the highly $T_c$ enhancement in FeSe/STO system.

In closing, we note that the direct imaging of band bending of an interfacial 2DEG system by EELS at liquid helium temperatures may be of interest and valuable for the study of other interfacial systems.

**Method**

**MBE growth**. Thin film growth for transport measurement was performed using a custom-built ultrahigh vacuum (UHV) MBE system with a base pressure lower than $5\times10^{-10}$ Torr. The insulating STO(001) substrates were prepared with standard chemical and thermal treatments to obtain a uniform $TiO_x$-terminated surface[7]. Then, the heat-treated STO substrates were transferred into the UHV MBE chamber and annealed at 600℃ for 1h. FeSe films were grown by co-evaporating Fe(99.995%) from an E-gun cell and Se(99.999%) from a Knudsen cell with a flux ratio of 1:20 on the STO substrate at 330 °C. The Fe and Se concentration in the films were determined by their ratio obtained *in situ* during growth using separate quartz crystal monitors. The growth rate for the films was approximately 0.2 UC/min. Epitaxial growth was monitored by *in situ* reflective high-energy electron diffraction (RHEED), where the high crystal quality and the atomically flat surface were confirmed by the streaky and sharp "1×1" patterns (Fig. S1).

**Atomic imaging and EELS measurements.** High resolution scanning transmission electron microscopy (STEM) images were acquired using high-angle-annular-dark-field (HAADF) detector on a double aberration correction JOEL ARM-200CF transmission electron microscope at Brookhaven National Lab. Atomically resolved EELS spectrum images were taken under STEM mode, at $T$=10K and room temperature, respectively, with the Gatan GIF Quantum ER dual EELS spectrometer. The low-temperature experiments were carried out using a custom-designed, low-drift liquid-He stage newly developed by Gatan for atomic imaging and spectroscopy. The temperature is monitored by a factory-calibrated silicone diode. The spatial increment of the line scan of the spectrum images was ~0.02nm. Both low-loss, including the zero-loss peak, and core-loss spectra were collected simultaneously for precise calibration to obtain the precise energy shift.

**Transport measurements.** The transport measurements were performed *ex-situ* on the FeSe/STO thin films. To avoid possible contamination, a 14UC thick epitaxial FeTe capping layer and another 10nm Te layer were deposited on top of the FeSe films before taken out of the growth chamber for transport measurements. The Hall and longitudinal resistances were measured using a commercial Quantum Design Physical Property Measurement System (PPMS) (1.8K, 9T) with the excitation current flowing in the film plane and the magnetic field applied perpendicular to the plane. The backgate voltage was applied using the Keithley 2450 from -200V to +200V.

**References and notes**

1    Song, Y. J. *et al.* Superconducting Properties of a Stoichiometric FeSe Compound and Two Anomalous Features in the Normal State. *Journal of the Korean Physical Society* **59**, 312-316, doi:10.3938/jkps.59.312 (2011).

2    Wang, Q. Y. *et al.* Interface-Induced High-Temperature Superconductivity in Single Unit-Cell FeSe Films on SrTiO3. *Chinese Physics Letters* **29**, 037402, doi:10.1088/0256-307x/29/3/037402 (2012).

3    Liu, D. F. *et al.* Electronic origin of high-temperature superconductivity in single-layer FeSe superconductor. *Nat. Commun.* **3**, 931, doi:10.1038/ncomms1946 (2012).

4    He, S. L. *et al.* Phase diagram and electronic indication of high-temperature superconductivity at 65 K in single-layer FeSe films. *Nat. Mater.* **12**, 605-610, doi:10.1038/nmat3648 (2013).

5    Tan, S. Y. *et al.* Interface-induced superconductivity and strain-dependent spin density waves in FeSe/SrTiO3 thin films. *Nat. Mater.* **12**, 634-640, doi:10.1038/nmat3654 (2013).

6   Lee, J. J. *et al.* Interfacial mode coupling as the origin of the enhancement of T-c in FeSe films on SrTiO3. *Nature* **515**, 245-248, doi:10.1038/nature13894 (2014).

7   Zhang, W. H. *et al.* Direct Observation of High-Temperature Superconductivity in One-Unit-Cell FeSe Films. *Chinese Physics Letters* **31**, 017401 doi:10.1088/0256-307x/31/1/017401 (2014).

8   Zhang, W. H. *et al.* Interface charge doping effects on superconductivity of single-unit-cell FeSe films on SrTiO3 substrates. *Phys. Rev. B* **89**, 060506(R), doi:10.1103/PhysRevB.89.060506 (2014).

9   Liu, X. *et al.* Dichotomy of the electronic structure and superconductivity between single-layer and double-layer FeSe/SrTiO3 films. *Nat. Commun.* **5**, 5047, doi:10.1038/ncomms6047 (2014).

10  Peng, R. *et al.* Tuning the band structure and superconductivity in single-layer FeSe by interface engineering. *Nat. Commun.* **5**, 5044, doi:10.1038/ncomms6044 (2014).

11  Sun, Y. *et al.* High temperature superconducting FeSe films on SrTiO3 substrates. *Scientific Reports* **4**, 6040 doi:10.1038/srep06040 (2014).

12  Deng, L. Z. *et al.* Meissner and mesoscopic superconducting states in 1-4 unit-cell FeSe films. *Phys. Rev. B* **90**, 214513, doi:10.1103/PhysRevB.90.214513 (2014).

13  Ge, J. F. *et al.* Superconductivity above 100 K in single-layer FeSe films on doped SrTiO3. *Nat. Mater.* **14**, 285-289 (2015).

14  Fan, Q. *et al.* Plain s-wave superconductivity in single-layer Fe Se on SrTiO3 probed by scanning tunnelling microscopy. *Nature Physics* **11**, 946-952, doi:10.1038/nphys3450 (2015).

15  Huang, D. *et al.* Revealing the Empty-State Electronic Structure of Single-Unit-Cell FeSe/SrTiO3. *Physical Review Letters* **115**, 017002, doi:10.1103/PhysRevLett.115.017002 (2015).

16  Li, F. S. *et al.* Atomically resolved FeSe/SrTiO3(001) interface structure by scanning transmission electron microscopy. *2D Mater.* **3**, 024002, doi:10.1088/2053-1583/3/2/024002 (2016).

17  Zhao, W. W., Chang, C. Z., Xi, X. X., Mak, K. F. & Moodera, J. S. Vortex phase transitions in monolayer FeSe film on SrTiO3. *2D Mater.* **3**, 024006, doi:10.1088/2053-1583/3/2/024006 (2016).

18  Zhang, S., et al. The Role of SrTiO 3 Phonon Penetrating into thin FeSe Films in the Enhancement of Superconductivity. *arXiv:1605.06941v2* (2016).

19  Xiang, Y. Y., Wang, F., Wang, D., Wang, Q. H. & Lee, D. H. High-temperature superconductivity at the FeSe/SrTiO3 interface. *Phys. Rev. B* **86**, 134508, doi:10.1103/PhysRevB.86.134508 (2012).

20  Liu, K., Lu, Z. Y. & Xiang, T. Atomic and electronic structures of FeSe monolayer and bilayer thin films on SrTiO3 (001): First-principles study. *Phys. Rev. B* **85**, 235123, doi:10.1103/PhysRevB.85.235123 (2012).

21  Bang, J. *et al.* Atomic and electronic structures of single-layer FeSe on SrTiO3(001): The role of oxygen deficiency. *Phys. Rev. B* **87**, 220503(R), doi:10.1103/PhysRevB.87.220503 (2013).

22  Li, B., Xing, Z. W., Huang, G. Q. & Xing, D. Y. Electron-phonon coupling enhanced by the FeSe/SrTiO3 interface. *J. Appl. Phys.* **115**, 193907, doi:10.1063/1.4876750 (2014).

23   Zhou, Y. & Millis, A. J. Charge transfer and electron-phonon coupling in monolayer FeSe on Nb doped SrTiO3. *arXiv:1603.02728v1* (2016).

24   Miyata, Y., Nakayama, K., Sugawara, K., Sato, T. & Takahashi, T. High-temperature superconductivity in potassium-coated multilayer FeSe thin films. *Nat. Mater.* **14**, 775-779, doi:10.1038/nmat4302 (2015).

25   Song, C. L. *et al.* Observation of Double-Dome Superconductivity in Potassium-Doped FeSe Thin Films. *Physical Review Letters* **116**, 157001, doi:10.1103/PhysRevLett.116.157001 (2016).

26   J. Shiogai, Y. I., T. Mitsuhashi, T. Nojima, A. Tsukazaki. Electric-field-induced superconductivity in electrochemically etched ultrathin FeSe films on SrTiO3 and MgO. *Nature Physics* **12**, 42-46 (2016).

27   Lei, B. *et al.* Evolution of High-Temperature Superconductivity from a Low-T-c Phase Tuned by Carrier Concentration in FeSe Thin Flakes. *Physical Review Letters* **116**, 077002, doi:10.1103/PhysRevLett.116.077002 (2016).

28   Tang, C. J. *et al.* Superconductivity dichotomy in K-coated single and double unit cell FeSe films on SrTiO3. *Phys. Rev. B* **92**, 180507(R), doi:10.1103/PhysRevB.92.180507 (2015).

29   Tamai, A. *et al.* Strong Electron Correlations in the Normal State of the Iron-Based FeSe0.42Te0.58 Superconductor Observed by Angle-Resolved Photoemission Spectroscopy. *Physical Review Letters* **104**, 097002, doi:10.1103/PhysRevLett.104.097002 (2010).

30  Yin, Z. P., Haule, K. & Kotliar, G. Kinetic frustration and the nature of the magnetic and paramagnetic states in iron pnictides and iron chalcogenides. *Nat. Mater.* **10**, 932-935, doi:10.1038/nmat3120 (2011).

31  He, J. F. *et al.* Electronic evidence of an insulator-superconductor crossover in single-layer FeSe/SrTiO3 films. *Proceedings of the National Academy of Sciences of the United States of America* **111**, 18501-18506, doi:10.1073/pnas.1414094112 (2014).

32  Hu, H. F. *et al.* Impact of interstitial oxygen on the electronic and magnetic structure in superconducting $Fe_{1+y}TeO_x$ thin films. *Phys. Rev. B* **90**, 5, doi:10.1103/PhysRevB.90.180504 (2014).

33  The program FEFF is an ab initio multiple-scattering code for calculating excitation spectra and electronic structure, based on a real space Green's function approach including a screened core-hole, inelastic losses and self-energy shifts, and Debye-Waller factors. The calculation performed in this code can directly give information about relativistic electron energy loss spectroscopy.

34  Rehr, J. J. *et al.* Ab initio theory and calculations of X-ray spectra. *C. R. Phys.* **10**, 548-559, doi:10.1016/j.crhy.2008.08.004 (2009).

35  Rehr, J. J., Kas, J. J., Vila, F. D., Prange, M. P. & Jorissen, K. Parameter-free calculations of X-ray spectra with FEFF9. *Phys. Chem. Chem. Phys.* **12**, 5503-5513, doi:10.1039/b926434e (2010).

36  Li, M. D. *et al.* Experimental Verification of the Van Vleck Nature of Long-Range Ferromagnetic Order in the Vanadium-Doped Three-Dimensional

Topological Insulator Sb2Te3. *Physical Review Letters* **114**, 146802, doi:10.1103/PhysRevLett.114.146802 (2015).

37  Santander-Syro, A. F. *et al.* Two-dimensional electron gas with universal subbands at the surface of SrTiO3. *Nature* **469**, 189-193, doi:10.1038/nature09720 (2011).

38  Meevasana, W. *et al.* Creation and control of a two-dimensional electron liquid at the bare SrTiO3 surface. *Nat. Mater.* **10**, 114-118, doi:10.1038/nmat2943 (2011).

39  Bel, R., Behnia, K. & Berger, H. Ambipolar Nernst effect in NbSe2. *Physical Review Letters* **91**, 066602, doi:10.1103/PhysRevLett.91.066602 (2003).

40  Seo, J. J. *et al.* Superconductivity below 20 K in heavily electron-doped surface layer of FeSe bulk crystal. *Nat. Commun.* **7**, 11116, doi:10.1038/ncomms11116 (2016).

**Acknowledgment**

We thank Jainendra Jain, DungHai Lee, Kin Fai Mak, Bangzhi Liu, Xiaoxiang Xi, Ludi Miao, Qingze Wang for fruitful discussions. This work is supported by the Penn State MRSEC, funded by the NSF under grants DMR 1420620. Work at BNL was supported by the US-DOE-BES, Materials Science and Engineering Division, under contract No. DE-SC0012704. Work at MIT was supported by grants NSF (DMR-1207469), NSF (DMR-0819762) (MIT MRSEC), ONR (N00014-13-1-0301), and the STC Center for Integrated Quantum Materials under NSF grant DMR-1231319.

**Author contributions**

W.Z., M.L., C.Z.C and M.H.W.C. conceived and designed the research along with Y.Z.. W.Z. made the devices and performed the transport measurements with the help of J.J. and M.H.W.C.. C.Z.C. grew the FeSe/STO films with the help of J.S.M.. M.L. and L.W. performed the EELS measurements with the help of Y.Z.. W.Z., M.L., C.Z.C, C.L., J. S. M and M.H.W.C. together analyzed and explained the data. W.Z. and M.H.W.C. wrote the manuscript with contributions from all authors.

**Additional information**

Supplementary information is available in the online version of the paper. Reprints and permissions information is available at www.nature.com/reprints.

Correspondences and requests for materials should be addressed to M.H.W.C., Y.Z. or C.Z.C.

**Competing financial interests**

The authors declare no competing financial interests.

Figures. 1-4

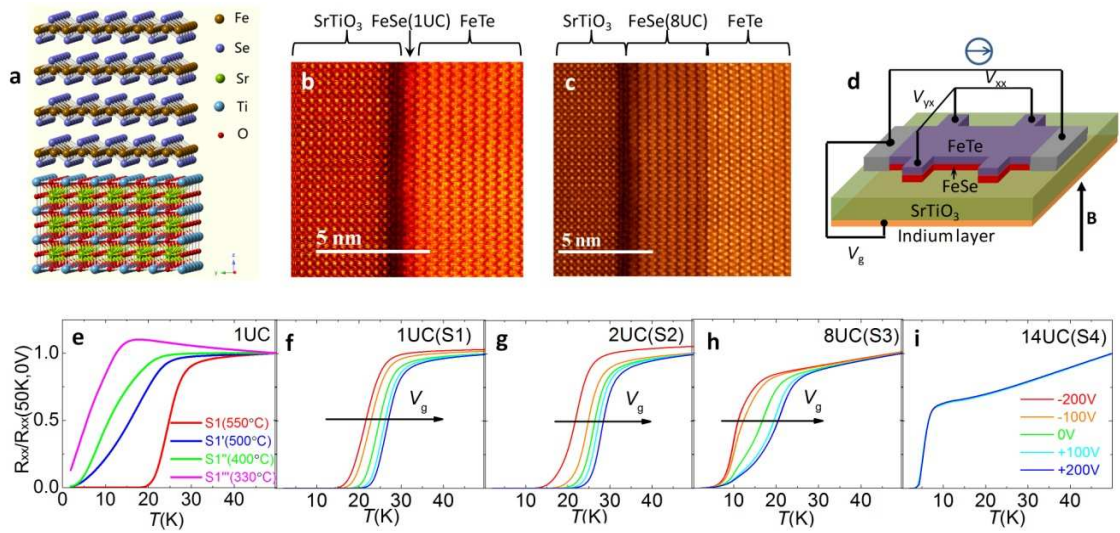

**Figure 1| Superconducting FeSe films on STO substrates.** (a) Super-cell of FeSe on top of STO(001). (b-c) High-angle-annular-dark-field scanning transmission electron microscopy (HAADF-STEM) image of the FeSe films with thickness of 1UC(b), 8UC(c) on STO substrates with 14UC FeTe capping layer on top. (d) Schematics of the gate tuned six-terminal Hall-bar device of FeSe films on STO with FeTe capping layer. (e) Normalized $R_{xx}$ vs. $T$ for 1UC films under different annealing temperatures for 2h during growth: S1(550°C), S1'(500°C), S1''(400°C) and S1'''(330°C). (f-i) Normalized $R_{xx}$ vs. $T$ at various backgating voltage $V_g$ for the 1UC(S1)(f), 2UC(S2)(g), 8UC(S3)(h) and 14UC(S4)(i) films under the optimal annealing condition at 550 °C for 2h. All transport results shown in Figs. 1e to 1i were measured under zero magnetic field.

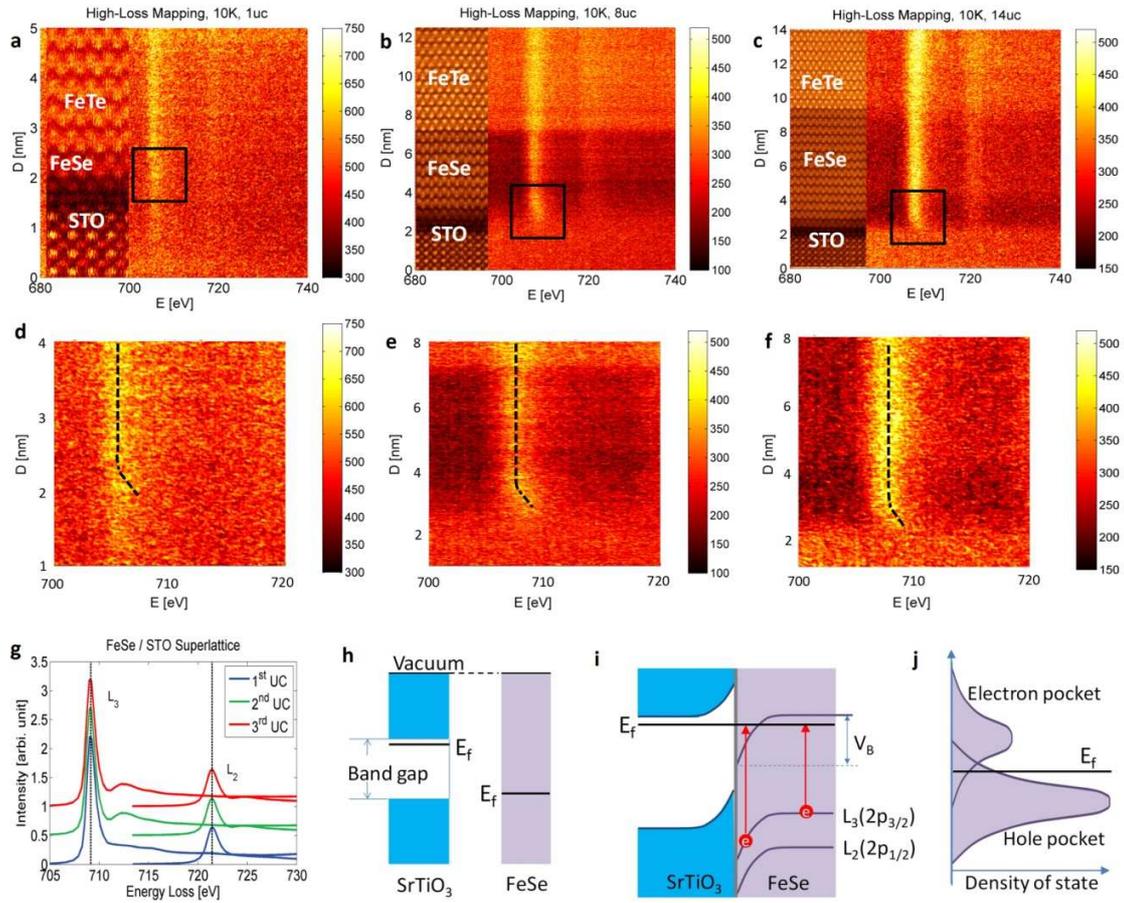

**Figure 2| Atomically resolved STEM-EELS (electron energy-loss spectroscopy) results for 1UC(S1), 8UC(S3), 14UC(S4) films at 10K.** (a-c) Core-loss EELS mapping with an energy range between 680 and 740eV for 1UC(a), 8UC(b) and 14UC(c) films. Blue-shift of Fe's $L_3$ peaks are observed in FeSe films that is within 2UC from the STO substrate in all 3 samples. (d-f) The zoom-in images at the squares shown in (a-c), respectively. The dash lines are guided by eyes. (g) FEFF simulation of the core-loss EELS spectra using the super-cell in Fig.1a. (h) Schematic of work function difference between STO and FeSe. (i) Schematic depiction of band bending in the FeSe region induced by electron transfer from the proximal STO interface. Due to band bending, Fe's $L_3$ ($2p_{3/2}$) and $L_2$ ($2p_{1/2}$) levels close to the interface bend accordingly, giving a blue-shift of the electron energy loss. $V_B$ is total potential variation. (j) At the interface, density of states of hole pocket is higher than that of electron pocket. It needs more electrons to fill up hole pocket as compared with electron pocket for the same $E_F$ shift.

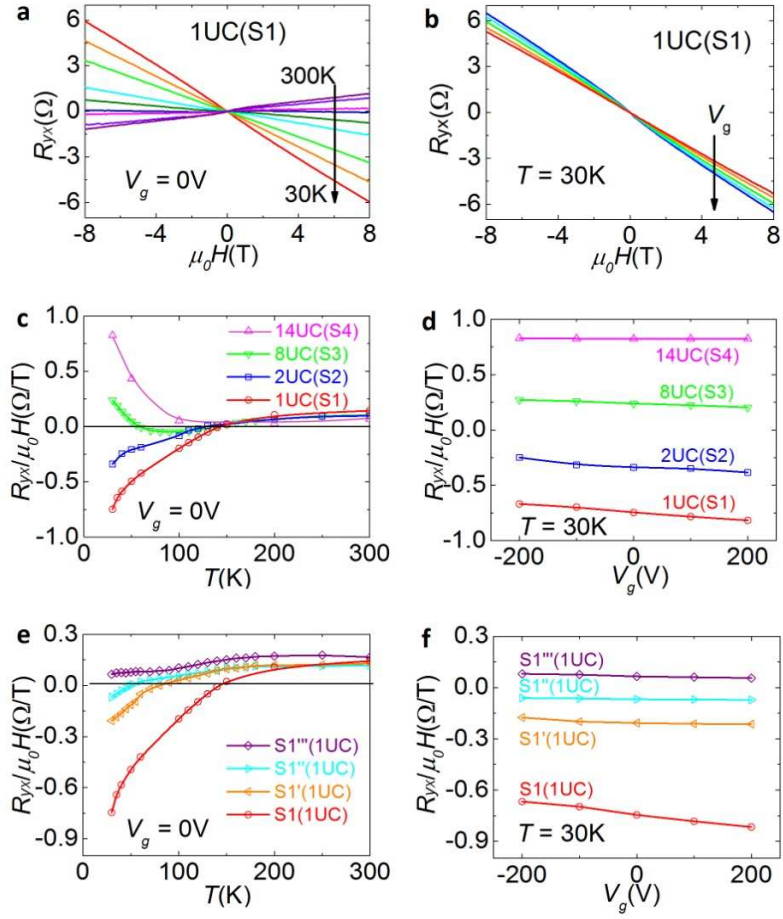

**Figure 3| Hall transport results for seven FeSe/STO samples studied.** (a) $R_{yx}$ vs. $\mu_0H$ at $V_g$= 0V at different temperatures from 300K to 30K for the 1UC(S1) film. (b) $R_{yx}$ vs. $\mu_0H$ at $T$=30K at different gate voltages ranging from -200V to +200V for the 1UC(S1) film. The black arrow indicates the direction of increasing $V_g$. (c) $R_H$ as function of $T$ at $V_g$=0V for the 1UC(S1), 2UC(S2), 8UC(S3), 14UC(S4) films. (d) $R_H$ as function of $V_g$ at $T$=30K for the 1UC(S1), 2UC(S2), 8UC(S3), 14UC(S4) films. (e) $R_H$ as function of $T$ under $V_g$=0V for 1UC films S1, S1', S1'', S1''' annealed at different temperatures. (f) $R_H$ as a function of $V_g$ at $T$=30K for 1UC films S1, S1', S1'', S1'''.

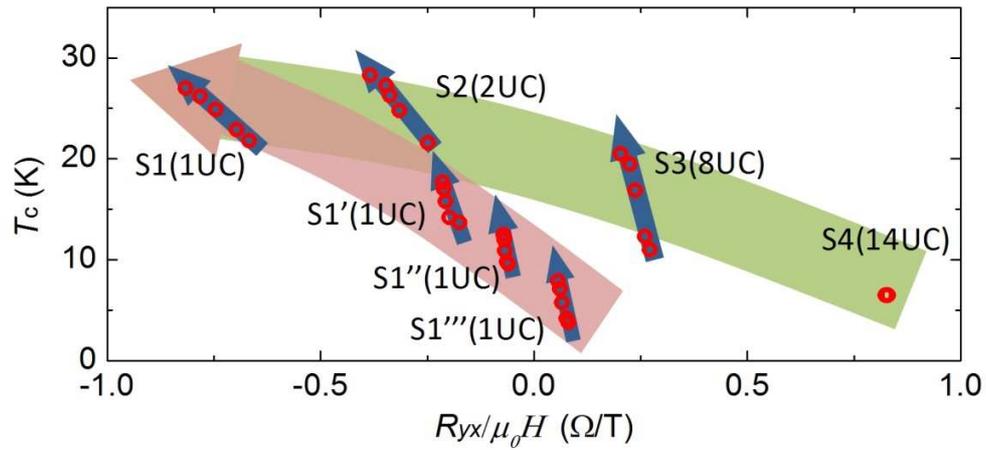

**Figure 4| The effect of backgating, annealing temperature and film thickness variation on the superconducting transition temperature $T_c$ and Hall coefficient $R_H$ at 30K for the 7 FeSe/STO samples studied in this experiment**. The 7 samples are respresented by the blue arrows. The red circles show $T_c$ and $R_H$ values, from bottom to top, at backgate voltages of -200, -100, 0, 100 and 200 V (position of red circles deduced from Fig. 1& Fig. 3 and Section VI &VII in Supplementary Information). The broad pink arrow groups the 1UC films annealed at different temperatures and the green arrow groups samples with different thicknesses annealed at 550℃. This figure shows that backgating is particularly effective in enhancing $T_c$ for thin FeSe films with minimal effect in $R_H$.